\documentclass[a4paper,11pt]{article}
\usepackage{pos}
\usepackage{subcaption}
\usepackage{hyperref}
\usepackage{lineno}

\title{A Posterior Analysis on IceCube Double Pulse Tau Neutrino Candidates}

\author{The IceCube Collaboration \\{\normalsize \normalfont(a complete list of authors can be found at the end of the proceedings)}}

\emailAdd{tianwei1997@sjtu.edu.cn}

\abstract{The IceCube Neutrino Observatory at the South Pole detects Cherenkov light emitted by charged secondary particles created by primary neutrino interactions. Double pulse waveforms can arise from charged current interactions of astrophysical tau neutrinos with nucleons in the ice and the subsequent decay of tau leptons. The previous 8-year tau double pulse analysis found three tau neutrino candidate events. Among them, the most promising one observed in 2014 is located very near the dust layer in the middle of the detector. A posterior analysis on this event will be presented in this paper, using a new ice model treatment with continuously varying nuisance parameters to do the targeted Monte Carlo re-simulation for tau and other background neutrino ensembles. The impact of different ice models on the expected signal and background statistics will also be discussed.

\vspace{4mm}
{\bfseries Corresponding authors:}
Wei Tian$^{1*}$,Fuyudi Zhang$^{1}$, Donglian Xu$^{1,2}$\\
{$^{1}$ \itshape Tsung-Dao Lee Institute, Shanghai Jiao Tong University, Shanghai, China}\\
{$^{2}$ \itshape School of Physics and Astronomy, Shanghai Jiao Tong University, Shanghai, China} \\

} % from abstract

\FullConference{37$^{\rm{th}}$ International Cosmic Ray Conference (ICRC 2021)\\
		July 12th -- 23rd, 2021\\
		Online -- Berlin, Germany}

\begin{document}
\maketitle

\clearpage

%----------------------------------------------------------------------------------------
%	Introduction
%----------------------------------------------------------------------------------------

\section{Introduction} \label{sec:intro}
The IceCube Neutrino Observatory is a cubic-kilometer scale Cherenkov telescope and instrumented at the South Pole at depths between 1450 and 2450 meters \cite{Aartsen:2016nxy}. The detector consists of 5160 digital optical modules (DOMs) arranged along 86 strings. Each DOM contains a 10-inch photomultiplier tube (PMT) to detect Cherenkov photons emitted by charged secondary particles produced in the primary neutrino interactions. High-energy astrophysical neutrinos may originate from hadronic processes in extreme astrophysical environments.  In spite of expected rare tau neutrino production at the sources, high-energy astrophysical neutrino fluxes for all flavors are expected to be approximately equal on Earth due to standard neutrino oscillations when travelling over cosmic distances. Therefore, the identification of tau neutrinos provides us an opportunity to constrain the neutrino production mechanisms at the sources and test the neutrino oscillation properties.
\par 
A previous double pulse tau neutrino analysis found three candidates with eight years of IceCube data \cite{Wille:2019pub}. The most promising one observed in 2014 occurred in the middle of the detector. In addition to waveform-based analyses, this event was identified as a tau neutrino candidate with the double cascade reconstruction method \cite{Stachurska:2019wfb} and machine learning \cite{Meier:2019ypu}. However, the main background for double pulse $\nu_{\tau}$ searches are $\nu_{\mu}$-induced tracks, in contrast to the double cascade method, in which the background is dominated by single cascades. Since the neutrino interaction position of the 2014 event is on top of the dust layer \cite{Ackermann:2006pva}, the ice might deplete Cherenkov light and distort the signals due to significantly more scattering and absorption than the surrounding ice. To explore this, a posterior analysis on this event was performed by using the target-volume re-simulation with a new treatment of ice model which propagates the systematic uncertainties. We start with recalling the double pulse algorithm developed in \cite{Wille:2019pub} and displaying the 2014 double pulse tau neutrino candidate event in section \ref{sec:double_pulse}. The dedicated re-simulation chain and settings are introduced in section \ref{sec:resimulation}. In section \ref{sec:analysis}, the posterior analyses based on re-simulation data will be presented. This work is summarized in section \ref{sec:summary}, along with a quick outlook for future work.

%----------------------------------------------------------------------------------------
%	Double Pulse Candidates
%----------------------------------------------------------------------------------------

\section{Double Pulse Tau Neutrino Candidates}\label{sec:double_pulse}

Cherenkov photons are captured by PMTs and create analog charge signals, which are digitized when exceeding the discriminator threshold of 0.25 photoelectrons. Analog Transient Waveform Digitizers (ATWDs) record 128 samples of the waveform for a total duration of $422\,\text{ns}$, corresponding to $3.3\,\text{ns}$ a bin. When a primary high energy tau neutrino undergoes a charged current (CC) interaction with nucleon, it will create a hadronic cascade and an outgoing tau lepton. The emitted tau propagates an average distance of $\sim50\,\text{m/PeV}$ and then can decay into electrons or hadrons with an inclusive branching ratio of $\sim$ $83\%$. Owing to this unique channel, if the $\nu_{\tau}$-CC interaction and subsequent tau decay happen favorably close to an optical sensor, the waveform in this sensor is expected to contain two resolvable peaks. Such a signature is a called double pulse waveform.

\begin{figure}
    \centering
    \begin{subfigure}[b]{0.38\linewidth}       
        \centering
        \includegraphics[width=\linewidth, height=38mm]{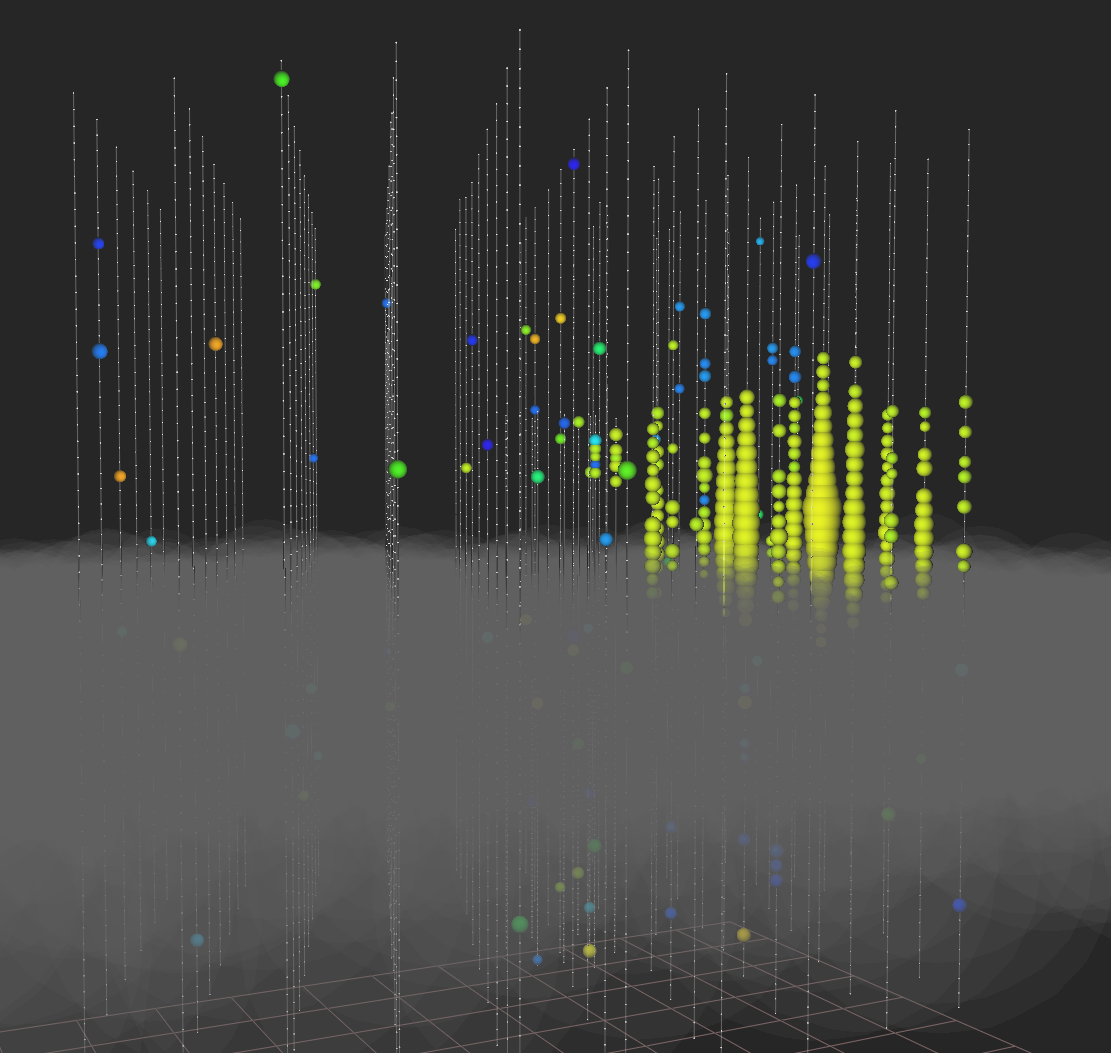}
        \caption{2014 event visualisation.}
        \label{fig:A}
    \end{subfigure}
    \hspace{3mm}
    \begin{subfigure}[b]{0.35\linewidth}       
        \centering
        \includegraphics[width=\linewidth]{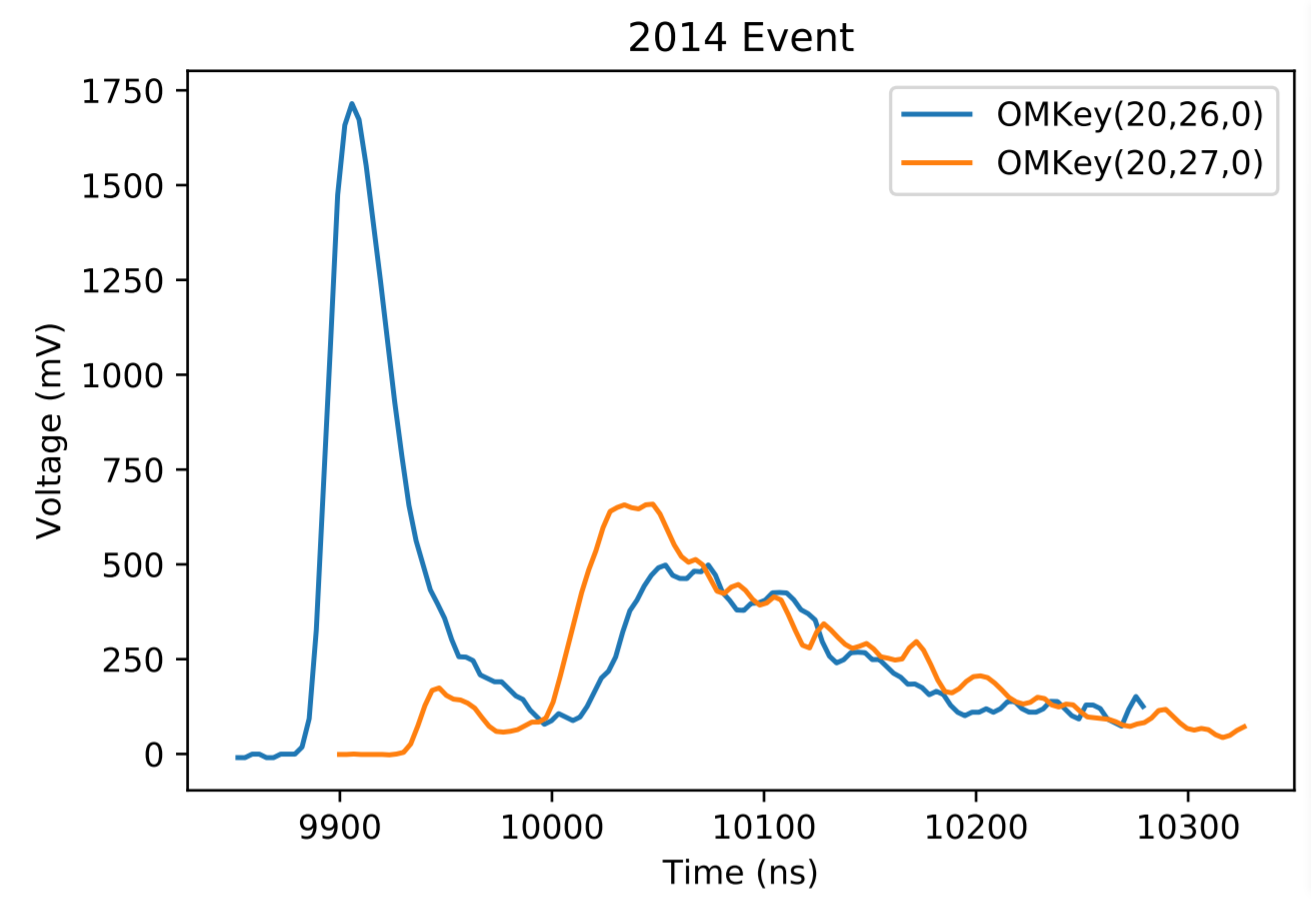}
        \caption{Two identified double pulse waveforms of 2014 event. }
        \label{fig:B}
    \end{subfigure}
    \caption{The event view and double pulse waveforms of the observed 2014 event. The right panel is taken from \cite{Wille:2019pub}, while the event view in the left panel is an data visualization, in which each dot or circle represents a DOM, the size of the circle is proportional to the amount of detected light, and the color corresponds to the relative light arrival time, with red earliest and blue latest, following the colors of the rainbow.}
    \label{fig:2014event}
\end{figure}

Two identified double pulse waveforms of the 2014 $\nu_{\tau}$ candidate \cite{Wille:2019pub} and its event view are depicted in Figure \ref{fig:2014event}. The reconstructed properties obtained by the maximum likelihood reconstruction algorithm \textsf{Monopod} \cite{Aartsen:2013vja} are listed in Table \ref{table:reco_params}. In terms of these properties, the tau neutrino with reconstructed energy of $\sim 93\,\text{TeV}$ could lead to a tau decay length of $\sim 10\,\text{m}$. It is noticeable that the reconstructed vertex of this candidate is on the top of the dust layer. The waveform shape maybe influenced by the local fluctuation of ice properties.
 
 To evaluate how likely this candidate is to be a tau neutrino, the targeted re-simulation was conducted with SnowStorm \cite{Aartsen:2019jcj}, a novel treatment of the optical property uncertainties of the ice with continuous variation of detector systematics. Taking advantage of target MC sets similar to the observed event, the extra perturbation contributed by background events can be analyzed. In addition, the impact of different ice models on the expected signal and background statistics will be discussed.

\begin{table}[ht]
\centering
\begin{tabular}{| l | r | r |}
    \hline
    Parameters &  \\ \hline \hline
    Energy &  93\,\text{TeV} \\ \hline
    Vertex Position & (309\,\text{m}, -205\,\text{m}, 63\,\text{m}) \\ \hline
    Zenith  & $\sim$\,54$\degree$ \\ \hline
    Azimuth  & $\sim$\,359$\degree$ \\ \hline
\end{tabular}
\caption{Reconstructed properties of the 2014 double pulse $\nu_{\tau}$ candidate event with \textsf{Monopod} \cite{Aartsen:2013vja}.}
\label{table:reco_params}
\end{table}

%----------------------------------------------------------------------------------------
%	Resimulation
%----------------------------------------------------------------------------------------

\section{Re-simulation Set up  of Double Pulse $\nu_{\tau}$ Candidate} \label{sec:resimulation}

The re-simulation chain is comprised of the following steps:

\textbullet\,\textbf{Event Generation} $\colon$ In this work, all primary neutrinos are generated with \textsf{LeptonInjector (LI)}~\cite{Abbasi:2020mwj}, a newly developed neutrino generator designed for large-volume Cherenkov neutrino telescopes such as IceCube. By using its volume-injection mode all three flavors of neutrinos are injected within a cylinder with radius of $25\,\text{m}$ and height of $50\,\text{m}$, whose center is the reconstructed vertex of the candidate. The dedicated LeptonInjector settings for this work are listed in Table \ref{table:generator}.

\textbullet\,\textbf{Propagation} $\colon$ Followed by neutrino generations, outgoing charged particles produced by the injected neutrinos are propagated with the \textsf{PROPOSAL} \cite{Koehne:2013gpa} and then are passed to \textsf{CLsim}~\cite{clsim} for photon propagation. The SnowStorm method is applied during photon propagation for modification of ice model parameters. So far, six SnowStorm parameters are implemented and the detailed settings are listed in Table \ref{table:ice_properties}. During photon propagation, each set of ice model parameters is sampled from Gaussian distribution for every 10 MC events. The IceWavePlusModes scales the ice absorption and scattering coefficients as depth depend using the icewave ice model \cite{Aartsen:2019jcj}. The unified HoleIce model depends on two parameters p0 and p1 which are implemented to change the acceptance probability of photons \cite{HoleIce}.

\begin{table}[ht]
\centering
\begin{tabular}{|c|c|}
\hline
\textbf{LI Parameters} & \textbf{Settings} \\
\hline
Flavors & $\nu_{e}$, $\nu_{\mu}$, $\nu_{\tau}$ \\
%Generated Energy & [50, 500]\,\text{TeV} for $\nu_{\tau}$, [50, 1000]\,\text{TeV} for $\nu_{\mu}$, [50, 250]\,\text{TeV} for $\nu_{e}$\\
Generated Energy & [50, 1000]\,\text{TeV} \\
Injected Center & (309\,\text{m}, -205\,\text{m}, 63\,\text{m}) \\
Injected Volume & radius = 25\,\text{m}, height = 50\,\text{m}\\
Zenith & [20 \degree, 80 \degree] \\
Azimuth & [0 \degree, 360 \degree] \\
\hline
\end{tabular}
\caption{Neutrino Generation Settings with the \textsf{LeptonInjetor} \cite{Abbasi:2020mwj}.}
\label{table:generator}
\end{table}

\begin{table}[ht]
\centering
\begin{tabular}{|c|c|c|}
\hline
\textbf{Ice Properties} &  \textbf{Sampling Distribution} & \textbf{Range} \\
\hline
IceWavePlusModes & Gaussian & Default\\
Absorption & Gaussian & $\mu$=1.0, $\sigma$=0.05 \\
Scattering & Gaussian & $\mu$=1.0, $\sigma$=0.05 \\
DOM Efficiency & Gaussian & $\mu$=1.0, $\sigma$=0.05 \\
Anisotropy & Gaussian & $\mu$=1.0, $\sigma$=0.1 \\
HoleIce Forward & Delta & [0.101569, -0.049344] \\
\hline
\end{tabular}
\caption{\textsf{SnowStorm} \cite{Aartsen:2019jcj} Parameter Settings.}
\label{table:ice_properties}
\end{table}

\textbullet\,\textbf{Detector} $\colon$ The next steps are the standard detector response simulation, Level1 and Level2 processing. 

\textbullet\,\textbf{Re-weighting} $\colon$ Events are re-weighted to the neutrino flux $E^{-2.5}$ via \textsf{LeptonWeighter} \cite{Abbasi:2020mwj}, the sister software of \textsf{LeptonInjector}.

\textbullet\,\textbf{Double Pulse Selection} $\colon$ All re-weighted events are passed to the double pulse algorithm (DPA), which was originally developed in \cite{Aartsen:2015dlt} and then extended in \cite{Wille:2019pub} as Local Coincidence DPA. The main idea behind this algorithm is to identify the rising and falling edge of the first pulse, which is followed by the second rising edge. The entire double pulse selection is presented in \cite{Wille:2019pub} in greater detail.

%----------------------------------------------------------------------------------------
%	Analysis
%---------------------------------------------------------------------------------------- 

\section{Results} \label{sec:analysis}
\subsection{Expected Event Rates in the Restricted Parameter Phase Space }
\begin{table}[ht]
\centering
\begin{tabular}{|c|c|c|c|c|c|c|}
\hline
\multirow{3}{1.2cm}{Neutrino Flavor}& \multicolumn{6}{c|}{Energy Groups (TeV)} \\
\cline{2-7} & \multicolumn{2}{c|}{[50,250]} & \multicolumn{2}{c|}{[250,500]} & \multicolumn{2}{c|}{[500,1000]}\\
\cline{2-7} & generated & passed & generated & passed & generated & passed \\
\hline
$\nu_{\tau}$ & $2 \times 10^4$ & 5785 & $2 \times 10^4$ & 29425 &  $ 2 \times 10^4$ & 41698 \\
$\nu_{\mu}$ & $2 \times 10^4$ & 161 & $2 \times 10^4$ & 1036 & $2 \times 10^4$ & 2137 \\
$\nu_{e}$ & $2 \times 10^4$ & 32 & $2 \times 10^4$ & 47& $2 \times 10^4$ & 53  \\
\hline
\end{tabular}
\caption{Number of generated MC events and number of MC events that passed the final level double pulse selection for each flavor neutrino. MC sets were generated in several energy groups. However, in the further study, we are only interested in [50, 500]\,\text{TeV} for $\nu_{\tau}$, [50, 1000]\,\text{TeV} for $\nu_{\mu}$ and [50, 250]\,\text{TeV} for $\nu_{e}$.}
\label{table:mc_events}
\end{table}

\begin{figure}[ht]
    \centering
    \includegraphics[width=0.95\textwidth]{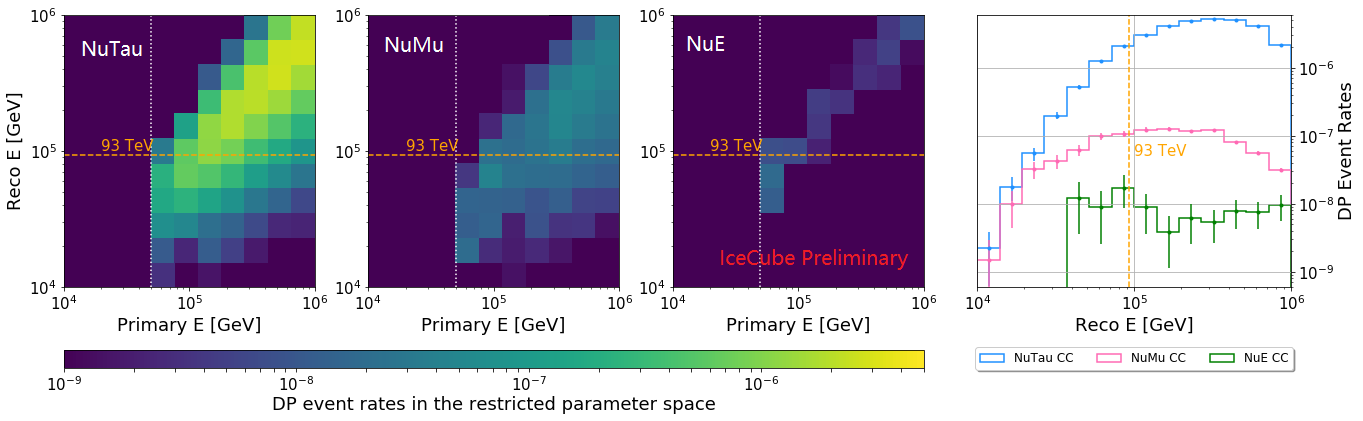}
    \caption{\textbf{Left three plots}: double pulse event rates as a function of MC primary neutrino energies and \textsf{Monopod} reconstructed energies per year in the restricted parameter space for each flavor neutrino. The horizontal line indicates 93\,\text{TeV}, the reconstructed energy of 2014 event. \textbf{Right plot}: distribution of reconstructed energy for each flavor neutrino.}
    \label{fig:distribution_energy}
\end{figure}

The re-simulation statistics are summarized in Table \ref{table:mc_events}. The number of generated MC events and the number of events that pass the final level double pulse selection are listed for each neutrino flavor and each energy group. Moreover, the event rates after re-weighting are depicted in Figure \ref{fig:distribution_energy} with respect to the energy. Three plots starting from the left indicate the correlated expected event rates as a function of primary neutrino energies and reconstructed energies for each flavor neutrino, while the plot on the right shows the independent distribution of reconstructed energy for tau, muon and electron neutrinos, respectively. The horizontal line indicates the reconstructed energy of the 2014 double pulse candidate, i.e. $93\,\text{TeV}$. 

It is clearly seen that the MC true energies and reconstructed energies show a good agreement for $\nu_{e}$, whereas such agreement smears out for $\nu_{\tau}$ and disappears for $\nu_{\mu}$. Due to the linear mapping between MC true energies and reconstructed energies, electron neutrinos with primary energies above $250\,\text{TeV}$ do not contribute to the double pulse events with reconstructed energy of $\sim 93\,\text{TeV}$. Compared with that, $\sim 40\%$ of $\nu_{\tau}$ double pulse events lying in the region around $93\,\text{TeV}$ are generated with primary energies between 100 and $200\,\text{TeV}$. The right panel of Figure \ref{fig:distribution_energy} shows that the distribution of $\nu_{\tau}$ (blue) continually rises up with increasing reconstructed energies and reaches the peak at $\sim 250\,\text{TeV}$, while the distribution of $\nu_{\mu}$ (pink) is relatively flat and reach a plateau between $\sim 70\,\text{TeV}$ and $\sim 350\,\text{TeV}$. Throughout almost the entire range of energies, the distribution of $\nu_{e}$ (green) fluctuates with large error bars due to limited statistics. As a consequence, the double pulse background is dominated by the $\nu_{\mu}$ induced events. Furthermore, at the energy of $\sim 93\,\text{TeV}$, i.e. the reconstructed energy of the 2014 candidate event, the $\nu_{\tau}$ induced double pulse signal event rate is an order of magnitude larger than the $\nu_{\mu}$ induced background event rate.

\begin{figure}[ht]
    \centering
    \includegraphics[width=0.75\textwidth,height=50mm]{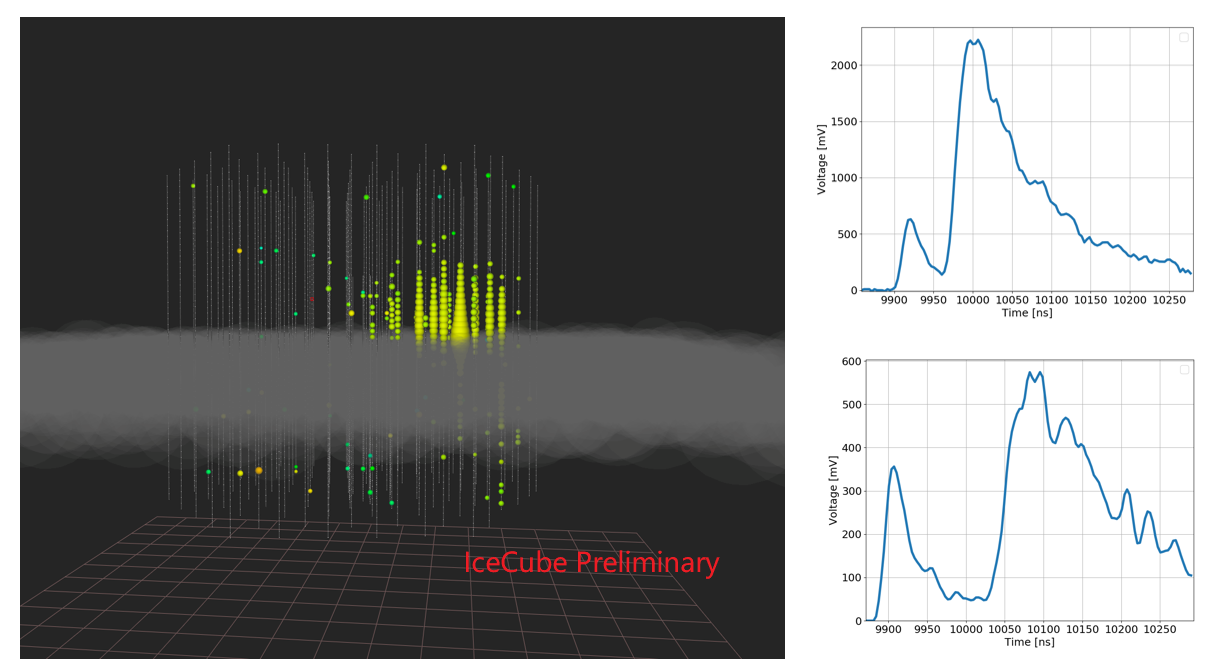}
    \caption{\textbf{Left}: Event view of the re-simulated muon-neutrino that passed the final double pulse selection. \textbf{Right}: two false positive double pulse waveforms of the $\nu_{\mu}$ event on the left.}
    \label{fig:numu_back}
\end{figure}

An example of a re-simulated $\nu_{\mu}$ event that passed the final level double pulse selection is sketched in Figure \ref{fig:numu_back}, along with two false positive double pulse waveforms. Here, the outgoing muon is hidden by the dust layer, resulting in a "cascade-like" event. The reconstructed energy of this event is about $90\,\text{TeV}$, which is close to the observed event.

\subsection{Double Pulse Purity and Impact of Systematic Uncertainties} 
\begin{figure}[ht]
    \centering
    \includegraphics[width=0.82\textwidth,height=4.6cm]{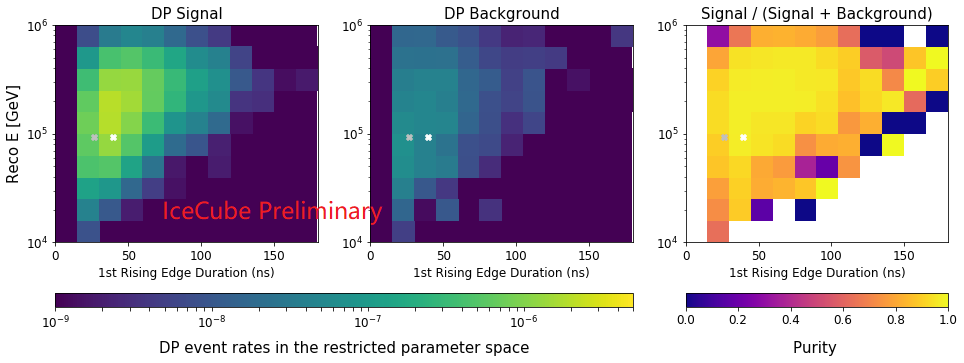}
    \caption{Double pulse event rates with respect to reconstructed energies and first rising edge durations for signal (\textbf{left}) and background (\textbf{middle}), respectively. \textbf{Right}: Purity as a function of reconstructed energy and the rising edge duration of the first pulse. Two selected double pulse waveforms of the 2014 event are indicated as two points.}
    \label{fig:sensitivity}
\end{figure}

The double pulse purity is defined as the tau neutrino induced double pulse signal event rate divided by the total selected double pulse event rate. Figure \ref{fig:sensitivity} shows the double pulse purity as a function of reconstructed energy and the rising edge duration of the first pulse in the waveform. In previous eight-year double pulse analyses, six features are used in the double pulse algorithm to select waveforms. They are steepness and duration of the first rising edge, first falling edge and second rising edge. The actual cuts are listed in Table 1 in \cite{Wille:2017kR}. In this work, the first rising edge duration is chosen to show the double pulse event rates and purity because the waveform with longer first rising edge duration tends to be a $\nu_{\tau}$-induced double pulse \cite{Wille:2019pub}. The left and middle panels of Figure \ref{fig:sensitivity} show the selected double pulse event rates as a function of the first rising edge durations and reconstructed energies for signal and background, respectively. Therefore, the ratio of signal event rate to the total event rate per bin is shown in the right panel, in the form of purity. The two points in each panel indicate two identified double pulse waveforms of the 2014 event. A time unit of $13.2\,\text{ns}$ is applied to calculate the time derivative of the waveform. Therefore, each bin along x-axis in this figure represents $\sim 13.2\,\text{ns}$ duration. It can be seen that the first rising edge durations of $\nu_{\tau}$ double pulse waveforms are usually longer than $26.4\,\text{ns}$, while $30\%$ of the double pulse background events exist with $26.4\,\text{ns}$ rising edge of the first pulse. This discrepancy leads to the lower purity in the region below $26.4\,\text{ns}$, in which one of the identified double pulse waveforms is located, as shown in the right panel. The right panel of figure shows that the purity of two bins in which two points lie are around $90\,\%$ and $97\,\%$, respectively.

\begin{figure}[ht]
    \centering
    \includegraphics[width=0.90\textwidth]{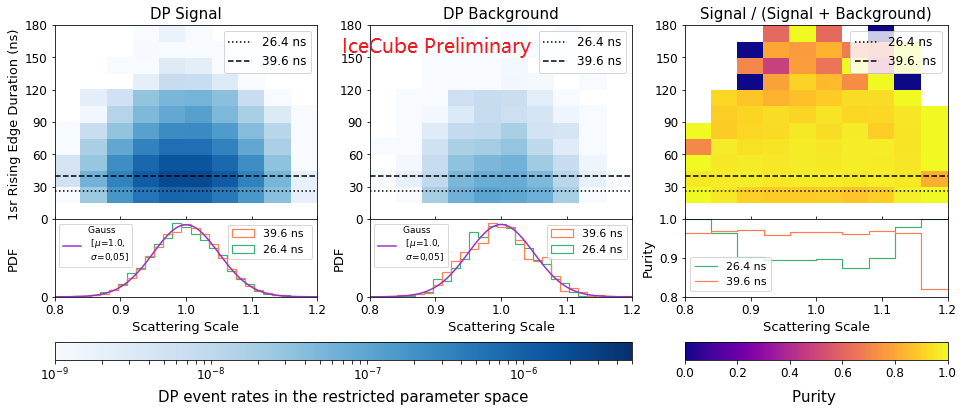}
    \caption{\textbf{Top}: Double Pulse signal (\textbf{left}) and background (\textbf{middle}) events rates and selection purity (\textbf{right}) as a function of scattering coefficient scale and rising edge duration of the first pulse. \textbf{Bottom}: Probability density distribution of double pulse signal (\textbf{left}) and background (\textbf{middle}) events and purity (\textbf{right}) as a function of scattering coefficient for events with $26.4\,\text{ns}$ (green) and $39.6\,\text{ns}$ (orange) long first rising edge duration. The Gaussian distribution used in SnowStorm for sampling the scattering coefficients is shown in purple.}
    \label{fig:scattering}
\end{figure}

Taking advantage of all inclusive detector systematics in a single SnowStorm MC ensemble, the correlated distributions of the first rising edge duration and ice scattering coefficient are shown in the top panel of Figure \ref{fig:scattering} for double pulse signal (left) and background (middle). The calculated purity is depicted on the right. Two horizontal lines in the top panel indicate the two first rising edge durations of the 2014 candidate with $26.4\,\text{ns}$ (dotted) and $39.6\,\text{ns}$ (dashed). The bottom left and middle subplots show the probability density distributions of double pulse events additionally selected with $26.4\,\text{ns}$ and $39.6\,\text{ns}$ first rising duration, while the bottom right shows the double pulse purity with respect to the scattering coefficients for aforementioned two group events. It is shown that the two PDF distributions almost follow the same Gaussian distribution used in SnowStorm settings for sampling the ice models, as shown in the bottom. In addition, for two sets of events that have the same first rising edge durations of the candidate, the selection purity remains almost unchanged and larger than $90\,\%$ within the sampling area of scattering coefficients, as shown in bottom right. Therefore, it can be concluded that the optical property uncertainties of ice models do not have significant impact on double pulse selection.

%----------------------------------------------------------------------------------------
%	Summary
%----------------------------------------------------------------------------------------

\section{Summary and Future Work} \label{sec:summary}

Based on the analysis of individual waveforms, the machine learning method \cite{Meier:2019ypu} found two tau neutrino candidates, while the local coincidence double pulse algorithm \cite{Wille:2019pub} found three tau neutrino candidates. The most promising one identified by both approaches happens on the top of the dust layer. In this work, the re-simulation chain was set up with SnowStorm to continuously vary detector systematics. The double pulse purity in the vicinity of the observed event is larger than $90\%$. Moreover, the variation of ice model parameters appears to have insignificant impact on double pulse selection. In future studies, atmospheric muons will be re-simulated as an additional component of background events to be considered. Additionally, the machine learning method \cite{Meier:2019ypu} will be combined into this work to complete the posterior analysis.

\bibliographystyle{ICRC}
\bibliography{references}

% Full authors list (ONLY FOR COLLABORATIONS)
\clearpage
\section*{Full Author List: IceCube Collaboration}

% \noindent \textbf{Note comment afterwards:} Collaborations have the possibility to provide an authors list in xml format which will be used while generating the DOI entries making the full authors list searchable in databases like Inspire HEP. For instructions please go to icrc2021.desy.de/proceedings or contact us under icrc2021proc@desy.de.\\

% \scriptsize
% \noindent
% first.author$^1$, 
% second.author$^2$, 
% third.author$^3$ % .... more names
% and 
% last.author$^{n}$ \\

% \noindent
% $^1$first.affiliation.
% $^2$second.affiliation. % .... more affiliation
% $^{m}$last.affiliation.

\scriptsize
\noindent
R. Abbasi$^{17}$,
M. Ackermann$^{59}$,
J. Adams$^{18}$,
J. A. Aguilar$^{12}$,
M. Ahlers$^{22}$,
M. Ahrens$^{50}$,
C. Alispach$^{28}$,
A. A. Alves Jr.$^{31}$,
N. M. Amin$^{42}$,
R. An$^{14}$
% ...
\\

\noindent
% ...
$^{12}$ Universit{\'e} Libre de Bruxelles, Science Faculty CP230, B-1050 Brussels, Belgium \\
% ...
$^{14}$ Department of Physics and Laboratory for Particle Physics and Cosmology, Harvard University, Cambridge, MA 02138, USA \\
% ...
$^{17}$ Department of Physics, Loyola University Chicago, Chicago, IL 60660, USA \\
$^{18}$ Dept. of Physics and Astronomy, University of Canterbury, Private Bag 4800, Christchurch, New Zealand \\
% ...
$^{22}$ Niels Bohr Institute, University of Copenhagen, DK-2100 Copenhagen, Denmark \\
% ...
$^{28}$ D{\'e}partement de physique nucl{\'e}aire et corpusculaire, Universit{\'e} de Gen{\`e}ve, CH-1211 Gen{\`e}ve, Switzerland \\
% ...
$^{31}$ Karlsruhe Institute of Technology, Institute for Astroparticle Physics, D-76021 Karlsruhe, Germany  \\
% ...
$^{42}$ Bartol Research Institute and Dept. of Physics and Astronomy, University of Delaware, Newark, DE 19716, USA \\
% ...
$^{50}$ Oskar Klein Centre and Dept. of Physics, Stockholm University, SE-10691 Stockholm, Sweden \\
% ...
$^{59}$ DESY, D-15738 Zeuthen, Germany \\
% ...

\subsection*{Acknowledgements}

\noindent
USA {\textendash} U.S. National Science Foundation-Office of Polar Programs,
U.S. National Science Foundation-Physics Division,
U.S. National Science Foundation-EPSCoR,
Wisconsin Alumni Research Foundation,
Center for High Throughput Computing (CHTC) at the University of Wisconsin{\textendash}Madison,
Open Science Grid (OSG),
Extreme Science and Engineering Discovery Environment (XSEDE),
Frontera computing project at the Texas Advanced Computing Center,
U.S. Department of Energy-National Energy Research Scientific Computing Center,
Particle astrophysics research computing center at the University of Maryland,
Institute for Cyber-Enabled Research at Michigan State University,
and Astroparticle physics computational facility at Marquette University;
Belgium {\textendash} Funds for Scientific Research (FRS-FNRS and FWO),
FWO Odysseus and Big Science programmes,
and Belgian Federal Science Policy Office (Belspo);
Germany {\textendash} Bundesministerium f{\"u}r Bildung und Forschung (BMBF),
Deutsche Forschungsgemeinschaft (DFG),
Helmholtz Alliance for Astroparticle Physics (HAP),
Initiative and Networking Fund of the Helmholtz Association,
Deutsches Elektronen Synchrotron (DESY),
and High Performance Computing cluster of the RWTH Aachen;
Sweden {\textendash} Swedish Research Council,
Swedish Polar Research Secretariat,
Swedish National Infrastructure for Computing (SNIC),
and Knut and Alice Wallenberg Foundation;
Australia {\textendash} Australian Research Council;
Canada {\textendash} Natural Sciences and Engineering Research Council of Canada,
Calcul Qu{\'e}bec, Compute Ontario, Canada Foundation for Innovation, WestGrid, and Compute Canada;
Denmark {\textendash} Villum Fonden and Carlsberg Foundation;
New Zealand {\textendash} Marsden Fund;
Japan {\textendash} Japan Society for Promotion of Science (JSPS)
and Institute for Global Prominent Research (IGPR) of Chiba University;
Korea {\textendash} National Research Foundation of Korea (NRF);
Switzerland {\textendash} Swiss National Science Foundation (SNSF);
United Kingdom {\textendash} Department of Physics, University of Oxford.

\end{document}